\documentclass[sigconf]{acmart}
\AtBeginDocument{%
  \providecommand\BibTeX{{%
    \normalfont B\kern-0.5em{\scshape i\kern-0.25em b}\kern-0.8em\TeX}}}

\copyrightyear{2021} 
\acmYear{2021} 
\setcopyright{acmcopyright}\acmConference[AIES '21]{Proceedings of the 2021 AAAI/ACM Conference on AI, Ethics, and Society}{May 19--21, 2021}{Virtual Event, USA}
\acmBooktitle{Proceedings of the 2021 AAAI/ACM Conference on AI, Ethics, and Society (AIES '21), May 19--21, 2021, Virtual Event, USA}
\acmPrice{15.00}
\acmDOI{10.1145/3461702.3462528}
\acmISBN{978-1-4503-8473-5/21/05}

\usepackage{tabularx}

\begin{document}

\title{Fairness and Data Protection Impact Assessments}

\author{Atoosa Kasirzadeh}
\affiliation{%
  \institution{University of Toronto \and Australian National University}
  \country{Australia}
}
\email{atoosa.kasirzadeh@anu.edu.au}

\author{Damian Clifford}
\affiliation{%
\institution{Australian National University}
  \country{Australia}
}
\email{damian.clifford@anu.edu.au}



\begin{abstract}
In this paper, we critically examine the effectiveness of the requirement to conduct a Data Protection Impact Assessment (DPIA) in Article 35 of the General Data Protection Regulation (GDPR) in light of fairness metrics. Through this analysis, we explore the role of the fairness principle as introduced in Article 5(1)(a) and its multifaceted interpretation in the obligation to conduct a DPIA. Our paper argues that although there is a significant theoretical role for the considerations of fairness in the DPIA process, an analysis of the various guidance documents issued by data protection authorities on the obligation to conduct a DPIA reveals that they rarely mention the fairness principle in practice. Our analysis questions this omission, and assesses the capacity of fairness metrics to be truly operationalized within DPIAs. We conclude by exploring the practical effectiveness of DPIA with particular reference to (1) technical challenges that have an impact on the usefulness of DPIAs irrespective of a controller’s willingness to actively engage in the process, (2) the context dependent nature of the fairness principle, and (3) the key role played by data controllers in the determination of what is fair.
\end{abstract} 




\maketitle

\section{Introduction}

The employment and deployment of machine learning algorithms in social contexts is widespread. These algorithms which are trained on massive amounts of data learn, without being programmed with special rules and principles, to predict with respect to a particular task (e.g., classification) about unobserved data. Machine learning algorithms impact various aspects of our private and public life by being embedded into socio-technical environments in areas as diverse as facial recognition \cite{buolamwini2018gender}, allocation of scarce medical goods \cite{obermeyer2019dissecting}, credit scoring \cite{khandani2010consumer}, and criminal justice decision making \cite{christin2015courts,julia2016machine}.

With a growing public awareness of their increasing impact, mitigating the unintended consequences of these algorithms for high-stake decision making has become the focus of much discussion in academic, business, and policy circles. One of the most popular ethical solutions to the mitigation of these unintended consequences is the operationalization of various fairness metrics in machine learning ecosystems. Indeed, in the academic and business circles, there has been a growing literature under the umbrella term of “fair machine learning” which claims to positively accommodate, via fairness metrics, some of the negative and/or unintended consequences of unfair prediction-based decision making \cite{friedler2019comparative,mehrabi2019survey}.

In policy making and the legal literature, there has been an ongoing discussion regarding the need for legislative reforms, also reflecting the need to cater for the negative impacts of algorithmic decision making \cite{yeung2018study}. Authors such as Nemitz (2018) have highlighted the importance of the General Data Protection Regulation (GDPR) in the regulation of algorithmic decision making \cite{nemitz2018constitutional}. In addition, some data protection authorities such as the Information Commissioner's Office (ICO) in the United Kingdom have emphasized the importance of the data protection framework in the policy guidance on developments in Artificial Intelligence \cite{ICO2017, ICO2020}. Through the lens of the GDPR, the ICO in particular appears to place significant weight on the importance of the fairness principle, as specified in Article 5(1)(a), to appropriately regulate machine learning systems \cite{ICO2017}. 

The fairness principle has been described as a core principle \cite{european2014privacy} and the cornerstones upon which the other principles contained in Article 5 of the GDPR are built \cite{bygrave2002data}. Despite its key importance, however, this principle remains somewhat of a nebulous concept and its relationship with the more accountability principle-orientated obligations on those processing personal data remains hard to characterize in a precise manner. 

The purpose of this paper is to assess the role of the fairness principle in the requirement to conduct a Data Protection Impact Assessment (DPIA) contained in Article 35 of the GDPR. If the fairness principle is to play a key role in mitigating the negative consequences of machine learning systems, it is certainly important to understand (1) how it is operationalized in the obligation to conduct DPIA and (2) how to evaluate the success of this mechanism in tackling the negative and/or unintended consequences of automated decision making. Building on this doctrinal legal analysis, we then explore how the fairness principle in the GDPR might be (or indeed, might not be) operationalized through fairness metrics. In addition, this paper critically examines the capacity of DPIAs to effectively accommodate the negative impacts of machine learning systems in light of insights gathered from the academic and business literature on the formalization and operationalization of various metrics of algorithmic fairness through the operation of the data protection fairness principle in the requirement to conduct a DPIA. This paper, therefore, takes some initial steps to connect these distinct bodies of literature.

The paper is structured as follows. In Section 2, we briefly introduce the GDPR and its key concepts and we examine the importance of the data protection fairness principle and the principle's role in the obligation to conduct a DPIA. Building on the insights gathered, in Section 3, we explore the interpretations of the data protection fairness principle in light of the pervasive literature on fairness metrics. Moreover, we investigate the potential roles that these interpretations can have in conducting a DPIA in Article 35 of the GDPR. We diagnose the pitfalls of using fairness metrics in light of the multiplicity of the interpretations of fairness and examine how fairness could play a more defined role, and how policy agendas distinct from data protection might effectively characterize what fairness means in specific contexts. The paper concludes in Section 4 by positioning the role of DPIAs and by calling for a more open debate regarding the role of the fairness principle and the intersection of data protection with different policy agendas in the determination of what processing operations should be deemed de facto unfair.

\section{The fairness principle and the requirement to conduct DPIA}

The GDPR aims to mitigate the power and information asymmetries between controllers (and processors) and data subjects or the natural persons to whom the personal data relates \cite{lynskey2014deconstructing}. The Regulation, as the key pillar of the European Union's data protection framework, formulates standards for the processing of personal data with personal data defined in Article 4(1) GDPR as,‘[...] any information relating to an identified or identifiable natural person (‘data subject’) [...]’. The Regulation affords rights to data subjects (e.g. erasure, access, rectification), imposes obligations on data controllers and processors, and assigns a monitoring role for data protection authorities. Data controllers are the natural or legal persons `which, alone or jointly with others, determine the purposes and means of the processing of personal data' (Article 4(7)), whereas the processor is the `natural or legal entity that processes personal data on behalf of the controller' (Article 4(8)). The  requirements that controllers and processors are subject to stem from the principles relating to the processing of personal data contained in Article 5 of the Regulation with these principles guiding the interpretation of the rights and obligations contained therein. The fairness principle, as stated in Articles 5(1)(a), alongside the lawfulness and transparency principles, is one of these key principles. 

The fairness principle is also mentioned specifically in Article 8(2) of the Charter of Fundamental Rights of the European Union, with Article 8 stipulating the right to data protection. Despite its fundamental role in debates about the preservation of human rights and artificial intelligence, however, the fairness principle has been largely unexplored and remains undefined in the data protection framework and case law. This is despite the fact that the requirement to process personal data ‘fairly’ is a standard-bearer in data protection. This in turn presents challenges for controllers in the fulfillment of their obligations. 

\subsection{Fairness, the fairness principle, and the GDPR}

As briefly mentioned in the introduction, the fairness principle is understood by many as the cornerstone upon which the other data protection principles are built. For instance, Bygrave observes that ‘it embraces and generates the other core principles of data protection laws’ \cite{bygrave2002data} and when positioned as such, fairness is connected to the protection against any negative consequences even in the absence of an intent to deceive on behalf of the controller when personal data are processed. When processing personal data, controllers are therefore obliged to consider the interests and reasonable expectations of the data subject. In a similar vein, \citet{malgieri2020concept} notes that `fairness refers to a substantial balancing of interests among data controllers and data subjects', and the principle is, therefore, effect-based in that `what is relevant is not the formal respect of procedures (in terms of transparency, lawfulness or accountability), but the substantial mitigation of unfair imbalances that create situations of ``vulnerability'''. This demonstrates the important connection between the fairness, lawfulness and transparency principles but also the accountability principle provided in Article 5(2) of the Regulation. 

Previous literature has explored the overlaps between the fairness, lawfulness and transparency principles in an attempt to delineate the precise role for fairness. \citet{clifford2018data}, for instance, divide the operation of the fairness principle into a process-oriented manifestation and an outcome-driven fair balancing. They argue that both run concurrently and inter-dependently throughout the application of the Regulation with respect to the ex ante and ex post rights and obligations contained in the GDPR. To clarify, the ex ante application of the fairness principle refers to the rights and obligations which apply prior to the processing of personal data such as the application of the conditions for lawful processing in Article 6(1) or the requirement to conduct a DPIA in Article 35; the ex post safeguards relate to the rights and obligations which apply during personal data processing and are most clearly manifested in the application of data subject rights. 

The more process- or procedure-orientated manifestation of the fairness principle results in the burdening of controllers with an obligation to be mindful of data subject's interests and capacities with reference to the ex ante and ex post operation of the information provision requirements. Therefore, the fairness principle is strongly connected to the transparency principle. The fair-balancing manifestation, on the other hand, refers to the weighing of the rights and interests of data subjects in determining the fairness of a processing operation in a more outcome-orientated manner, again with both ex ante and ex post manifestations. Interestingly, \citet{malgieri2020concept} observes that this dualist understanding of the role/manifestation of the fairness principle seems to have also been established in the modernized Council of Europe Convention 108. Irrespective of such a division, it is clear that the fairness principle is primarily concerned with mitigating the negative impacts of the power and information asymmetries between the controller (and processor) and data subject. 

Indeed, the division of the manifestation of the fairness principle could arguably be categorized within an overarching notion of the fairness principle as concerned with fair balancing in that the procedural fairness manifestations are also indicative of the need to take the rights and interests of the data subjects into account. This seems to align with the idea that fairness, at its core, refers to the need to prevent adverse effects and balance conflicting rights and interests. Here reference can be made for example to various guidance documents demonstrating the link between the fairness principle and non-discrimination \cite{EDPBDPbD, ICO2020}.

More broadly, the ICO has noted that fairness involves three elements: (1) a consideration of the effects on individuals, (2) the expectations of the data subject, and (3) the transparency of the data processing. Similarly, the French Data Protection Authority (CNIL) has stated that the fairness principle should be interpreted as a means of preventing unfair outcomes or impacts with the effects incorporating not only the perspective of the data subject but also a more collective one. It is important to note, however, that discriminatory effects are just one form that an unfair or unbalanced outcome may take. This appears to reflect the approach taken by the European Data Protection Board (EDPB) in its guidance on Data Protection by Design and by Default in which specific design elements are proposed as a means of considering the implementation of the fairness principle \cite{EDPBDPbD}. These design elements are listed in table 1.

\begin{table}
\centering
\begin{tabular} { | m{6.5em} | m{6.5cm}| }
 \hline
 Design element & EDPB explanation  \\ [0.5ex] 
 \hline\hline
 Autonomy & Data subjects should be granted the highest degree of autonomy possible to determine the use made of their personal data, as well as over the scope and conditions of that use or processing.  \\ 
  \hline
 Interaction & Data subjects must be able to communicate and exercise their rights in respect of the personal data processed by the controller.  \\
 \hline
 Expectation & Processing should correspond with data subjects’ reasonable expectations.  \\
 \hline
 Non-discrimination & The controller shall not unfairly discriminate against data subjects.  \\
 \hline
 Non-exploitation & The controller should not exploit   the needs or vulnerabilities of data subjects.  \\
 \hline
Consumer choice & The controller should not lock-in their users in an unfair manner. Whenever a   service processing personal data is proprietary, it   may create a lock-in to the service, which may not be fair, if it impairs the data subjects' possibility to exercise their right of data portability in accordance with Article 20. \\
\hline
Power balance & Power balance should be a key objective of the controller-data subject relationship. Power   imbalances should be avoided. When this is   not possible, they should be recognized and accounted for   with suitable countermeasures.\\
\hline
No risk transfer & Controllers should not transfer the risks of the enterprise to the data subjects.\\
\hline
No deception  & Data processing information and   options should be provided in an objective and neutral way, avoiding any deceptive or manipulative language or design.\\
\hline
Respect rights & The controller must respect the fundamental rights of data subjects and implement appropriate measures and safeguards and not impinge on those rights unless expressly justified by law.\\
 \hline
 Ethical & The controller should see the processing’s wider impact on individuals' rights and dignity.\\
 \hline
 Truthful & The controller must make available information about how they process personal data, they should act as they declare they will and not mislead the data subjects.\\
   \hline
Human intervention & The controller must incorporate   qualified human intervention that is capable of uncovering biases that machines may create in accordance with the right to not be subject to automated individual decision making in Article 22.\\
   \hline
 Fair algorithms   & Regularly   assess   whether   algorithms   are   functioning   in   line   with   the purposes and adjust the algorithms to mitigate uncovered biases and ensure   fairness   in   the   processing.   Data   subjects   should   be   informed about   the   functioning   of   the   processing   of   personal   data   based   on algorithms that analyze or make predictions about them, such as work performance,   economic   situation,   health,   personal   preferences,reliability or behavior, location or movements.\\ [1ex] 
 \hline
 \end{tabular}
 \caption{EDPB and the development of the fairness design elements}
\end{table}

This list of design elements aligns well with viewing fairness as a principle intended to counteract power and information asymmetries and thus as a protection against negative consequences stemming from personal data processing even in the absence of an intent to deceive on behalf of the controller. Indeed, some items on the list appear to be abstract examples of design elements that are unfair (Non-discrimination, Non-exploitation, Consumer choice, Power balance, No risk transfer, No deception and Truthful), whereas others seem to relate to specific counter-measures that may be used to prevent unfair outcomes (Interaction, Human intervention and Fair algorithms). The rest appear to reflect important underlying rights and values (Autonomy, Expectation, Respect rights, and Ethical). Thus, it seems uncontroversial to suggest that fairness is inherently linked with balancing competing rights and interests in an ecosystem dictated by power and information asymmetry. As a result, determining what is fair is couched in terms of balancing and analyzing the necessity and proportionality of the processing which feeds into the amorphous and context dependent nature of what might be deemed a fair outcome. 

\subsection{The Fairness principle and DPIAs}

In line with the above discussion, the requirement to conduct a DPIA and the preliminary assessment to determine whether a DPIA is required can be understood as examples of ex ante regulatory mechanism. Moreover, this requirement can be examined as `early warning systems' that aim to identify the impact of potential risks, and also to fairly balance and mitigate the potential risks with a clear connection to the accountability principle \cite{kosta20}. Indeed, according to Recital 84 `[t]he outcome of the assessment should be taken into account when determining the appropriate measures to be taken in order to demonstrate that the processing of personal data complies with this Regulation.' The data protection fairness and accountability principles go hand in hand with the controller responsible for the fair balancing of rights and interests when processing personal data. Fairness then manifests itself in the implementation of the rights and requirements provided by the framework to ensure a fair personal data processing ecosystem. 

Article 35(1) obliges controllers to perform a DPIA when a data processing operation (or set of similar operations), ‘is likely to result in a high risk to the rights and freedoms of natural persons’, and in particular if this operation makes use of new technologies. More specifically, Article 35(3) of the Regulation non-exclusively lists three specific cases when a DPIA is required, and Article 35(4) mandates the data protection authorities to publish a list of processing operations that are subject to the requirement to conduct a DPIA. Importantly, the key to determining what ‘fairly balanced’ personal data processing amounts to is to simply apply the checks and balances in the GDPR (i.e. including the obligation to conduct a DPIA). However, this does not eliminate the need to interpret what is ‘fair’ (or unfair) in the operation of the requirements through the lens of the fairness principle in Article 5(1)(a) of the Regulation. In other words, the Regulation and the requirement to conduct a DPIA are examples of the fair balance struck by the legislator between the competing rights and interests, but this does not eliminate the need to determine what the fairness principle as provided in Article 5(1)(a), as part of the balance struck by the legislator, means in a specific context. There is, therefore, a need for a better understanding of what fairness means through the lens of the requirement to conduct a DPIA and therefore, an exploration as to how this rather nebulous concept could be operationalized more effectively.

In her analysis of fundamental rights impact assessments in the context of automated decision making, \citet{janssen2020approach} includes the ‘balancing of risk and interests’ as one of her key benchmarks and frames this as something coming within the scope of the data protection fairness principle. However, despite the merits of understanding fairness as holding a key role in the requirement to conduct a DPIA, it remains unclear how fairness as a core principle of the Regulation actually applies to the requirement to conduct a DPIA. This is indicative of the fact that a review of the guidance literature exploring the role of the DPIA process reveals a very limited discussion of the fairness principle. For example, there is no reference to the fairness principle in the Article 29 Working Party guidance on the requirement to conduct a DPIA \cite{DPIA17}, the recent European Data Protection Survey's Report on the DPIAs conducted by the European Union's institutions \cite{ref10}, the CNIL and the Irish Data Protection Commissioner’s general guidance on the obligation to conduct a DPIA \cite{ref11}, the CNIL’s privacy impact assessment methodology \cite{ref12}, or the ICO's DPIA template \cite{ref13}. 

In contrast, some documents seem to make somewhat perfunctory references to the fairness principle. For instance, in its guidance on DPIAs, the ICO mentions that a DPIA may help demonstrate compliance with the `fairness and transparency requirements' \cite{ICO2021} and its guidance on artificial intelligence that a DPIA should include `an explanation of any relevant variation or margins of error in the performance of the system may affect the fairness of the personal data processing' \cite{ICO2020}. Given the lack of a more in-depth guidance, it is therefore necessary to analyze the role of the fairness principle in the requirement to conduct a DPIA in more detail.

More specifically, there are at least three places in which one can view a role for the fairness principle provided for in Article 5(1)(a) of the Regulation in Article 35. The first is in the determination of what is meant by ‘high risk’ in Article 35(1) where a failure to conduct a DPIA properly (or indeed at all) would seemingly breach the fairness principle. The second is in the interpretation of the situations identified as being specific cases of high risk in Article 35(3). Here reference can be made to Article 35(3)(a) which essentially makes a cross reference to the right not to be subject to an automated decision, including profiling contained in Article 22 of the Regulation. Indeed, it seems appropriate that the assessment of the impact of automated decision making should include a reference to the fairness of the processing operation in question given that such a decision may result in an unfair or biased outcome. As \citet{hacker2018teaching} notes, it would be odd to conclude that a model that racially discriminates processes personal data fairly. Therefore, this opens up the data protection toolbox to mitigate these challenges and therefore seemingly obliges controllers to consider the fairness of the processing in the DPIA. The third is, in a connected sense, the content of a DPIA including the items listed in Article 35(7) and for example, the assessment of the proportionality and necessity of the processing operations required under Article 35(7)(b).

Hence, although it remains somewhat implicit, it is certainly possible to plot the role of the fairness principle in the operation of the obligation to conduct a DPIA and the related provisions. What remains unclear, however, is how fairness should be effectively operationalized. The omission of a detailed analysis of the fairness principle from the various guidance documents issued by the data protection authorities listed above on the application of the requirement to conduct a DPIA is perhaps indicative of the process/procedural orientated ways that such accountability based requirements have been traditionally viewed. It seems, therefore, that generally speaking, as the starting point for the requirement to conduct a DPIA is strongly linked to the accountability principle, there is silence on the role of the fairness principle. This is despite the fact that fairness in Article 5(1)(a) must operate implicitly because it represents the need to fairly weigh the respective rights and interests at stake even in the operation of procedural rights and responsibilities. This omission is perhaps linked to the fact that a DPIA examines planned/future processing of personal data as opposed to ongoing processing operations (i.e. leaving the obligation to review to one side). Despite the lack of a specific and coherent analysis of the fairness principle in this DPIA documentation, there is clearly a role for the fairness principle in the DPIA process. It would be counter-intuitive and seemingly bizarre to disregard this ‘core’ principle in what is effectively an accountability and legitimacy check for future processing operations.

\section{Interpreting the fairness principle in light of fairness metrics}

The discussion so far suggests that there is  a need to think more deeply about the relationship between fairness as a ‘core’ principle of the GDPR and the methodological approach to a DPIA. Such an analysis would allow for a better understanding of the relationship between the principles provided for in Article 5 but also the malleable nature of the fairness principle. However, due to the broad role that fairness plays in the discussion, it is difficult to extract a concrete and actionable framework to assess fairness and inform a controller on how to undertake a balancing exercise given the complex context dependent nature of the principle in its operation. All the practical uncertainties of fairness in Article 5(1)(a) (i.e. in terms of the substantive outcome of a balancing exercise) effectively may mean that for a business, the only aspect that can really be entirely controlled is the conformity with the accountability-based process orientated requirements designed by the legislator to strike a fair balance between the competing rights and interests at stake. This statement will ring true where the underlying business model of the company (or a significant part of it) may in its entirety draw into question its compliance with the fairness principle. This is significant given the fact that guidance documentation on the application of the GDPR to developments in artificial intelligence (such as that issued by the ICO) seem to rely heavily on the fairness principle \cite{ICO2017, ICO2020}. 

The question thus becomes whether the literature on fair machine learning and fairness metrics could aid in formalizing the substantive content of what amounts to `fair' processing in relation to the obligation to conduct a DPIA to solidify the more substantive outcome-driven (i.e. fair balancing) role for the fairness principle in practice. This points to the need to bridge the gap between the high-level abstract formulation of the fairness principle in the GDPR and the literature on the variety of fairness metrics for making artificial intelligence systems fair. Indeed, given the growth of the literature on fairness metrics as developed by computer scientists \cite{hardt2016equality,dwork2012fairness,chouldechova2017fair,kusner2017counterfactual,kleinberg2018inherent,mehrabi2019survey,berk2018fairness,corbett2017algorithmic,friedler2019comparative,pleiss2017fairness}, fairness metrics are a prominent option for resolving this interpretational issue. We believe that this interpretation is conceptually necessary because the fairness principle is inherently linked to balancing rights and interests in a socio-technical ecosystem dictated by asymmetrical information power. To analyze whether this balancing has happened, some kind of concrete quantified metrics of fairness as a bridge requirement from the regulation frameworks to the artificial intelligence systems will provide important insight.

According to the literature on algorithmic fairness and fairness metrics, there are several approaches to formalize and quantify the conceptions of fairness. These approaches can be categorized, broadly, depending on whether we want to examine the notion of fairness in a statistical or an individualistic sense. Each of these senses can be interpreted in various ways. To provide a flavor of some of the possible notions of fairness metrics, we briefly review a small list of them. This list is by no means exhaustive. For a comprehensive survey, see \cite{friedler2019comparative,mehrabi2019survey}. Moreover, the appropriateness of the use of some of these metrics will depend on the kind of learning algorithm. For the purpose of this paper, however, we do not engage with this dependence. Our short review merely aims to provide some high level and basic insights in highlighting the challenges of this interpretation task. 

The simplest and perhaps the most straightforward conception of fairness is fairness through blindness to some sensitive attributes (e.g., race) that could be the basis of unfair treatment of the subject. In the context of machine learning systems, this means that in order to make the results of predictive algorithms fair, we need to make sure that these algorithms have no information about the sensitive attributes of the subject.\footnote{To understand how this idea might apply in practice, it is helpful to consider an algorithm that is used by a criminal justice system for making bail and parole decisions. This algorithm assigns a level of risk to each defendant. The risk assessment takes as input a set of individual's features such as their age and their previous offense history, and outputs the risk of the individual re-offending. Making the algorithm fair through blindness essentially means that any feature that prima facie is taken to be treated unfair should be removed, for instance, when a machine learning algorithm is being trained to evaluate the risk scores.} Unfortunately, this conception does not appear helpful in many applications in practice because very often (the conjunction of) some features such as demographic information, the postcode (in segregated cities), or the annual income operate as a good proxy for informing the algorithm about the sensitive attributes. This means that although no sensitive attribute is directly given as an input information to the algorithmic system, the collection of some non-sensitive attributes can reliably approximate some sensitive features to which the algorithm is supposed to be blind.

Another popular metric for the assessment of algorithmic fairness is statistical parity \cite{dwork2012fairness}. This characterization of fairness requires the predicted outcome of an algorithm to be statistically independent from the sensitive attributes. For instance, in the case of college admission, the predicted acceptance rates for both protected and unprotected groups should be the same (e.g. the acceptance rates of the applicants from different demographic groups must be equal). However, this notion of fairness might render misleading results, for instance when the underlying base rate for the protected and unprotected are different (e.g., fairness of arrest rate for violent crimes). Still another statistical fairness metric, equality of opportunity, measures whether those people who should qualify for an opportunity are equally likely to do so regardless of the group they are a member of \cite{hardt2016equality}. One limitation of this metric is as follows: if one of the goals of a fairness metric is to close the gap between the two subgroups, the metric will not help to achieve that goal. 

In addition to statistical fairness metrics, the conception of individual fairness aims to formalize and quantify the notion of fairness relative to the similar treatment of similar individuals \cite{dwork2012fairness}. However, measuring the similarity between two individuals in a metric space is an extremely tricky task. One way to provide a more concrete analysis of the notion of individual fairness is to use counterfactuals according to which a predictor's behavior must be compared across counterfactually similar individuals. For instance, \citet{kusner2017counterfactual} define a fair predictor to be the one that gives the same prediction had the individual were different with respect to some attributes, for example had the individual been of another race or gender. This demands an implicit assumption that everything else (except for the tweaked attributes) will be presumed the same for that individual. Unfortunately, this fairness metric also comes with some difficulties in analyzing and evaluating the counterfactual statements. See \citet{kasirzadeh2021use} for some principled arguments against the prevalent use of counterfactual fairness in social contexts.

The discussion so far only specifies some of the primary variants of fairness metrics of the more than twenty definitions discussed in the algorithmic fairness literature (each with their own benefits and conceptual flaws). However, we believe that this brief discussion is sufficient in allowing us to draw some general conclusions about the use of fairness metrics in interpreting the fairness principle. For instance, one of the most significant results in the literature on algorithmic fairness is the impossibility results \cite{chouldechova2017fair,kleinberg2018inherent}, which show that the simultaneous satisfaction of some of the desirable fairness metrics (except in some trivial cases) is mathematically impossible. Indeed, the abundance of the metrics for capturing algorithmic fairness gives us an important lesson, namely, that if we base the interpretation of the data protection fairness principle in the literature on algorithmic fairness metrics, data controllers will have a high degree of liberty in claiming a “fair” personal data processing ecosystem. This would result in a pluralistic interpretation of the principle on the basis of a codified interpretation of what fairness means and would therefore, seem destined to fail to really move beyond the abstract notion of fairness in the GDPR. 

Indeed, the move towards interpreting fairness metrics arguably belies the breadth of the fairness principle as provided for in the Regulation given that such metrics, for instance, seem focused on the removal of bias. Although non-discrimination can certainly be understood as a way in which the fairness principle plays a role in mitigating unfair outcomes, as discussed above, discriminatory effects are merely one example of an unfair outcome under the GDPR as opposed to the apparent equating of fairness and equality in the fairness metrics literature. As an example, in its guidance on Data Protection by Design and by Default, \citet{EDPBDPbD} states that
\begin{quote} ‘Fairness is an overarching principle which requires that personal data should not be processed in a way that is unjustifiably detrimental, unlawfully discriminatory, unexpected or misleading to the data subject. Measures and safeguards implementing the principle of fairness also support the rights and freedoms of data subjects, specifically the right to information (transparency), the right to intervene (access, erasure, data portability, rectify) and the right to limit the processing (right not to be subject to automated individual decision making and non-discrimination of data subjects in such processes).’\end{quote}
This aligns with viewing fairness as a principle designed to counteract information or power asymmetries and hence, as a protection against negative consequences stemming from personal data processing even in the absence of an intent to deceive on behalf of the controller, as presented above. But what does the broadness of this role for the principle mean in terms of its capacity to effectively cater for developments in artificial intelligence systems and algorithmic decision making? And what does this mean with respect to the requirement to  conduct a DPIA? 

Practically speaking, given that defining what is fair in terms of a substantive outcome is context dependent, it is suggested that performing the DPIA process in a thorough fashion in itself will often constitute a large part of doing what is ‘fair’. This point is indicative of the fact that, as mentioned earlier, a company’s business model may run counter to certain interpretations as to how the fairness principle plays a role in the interpretation of key rights and requirements and thus the categorisation of certain processing operations as unfair and unlawful. 

Of course, there are certainly clear instances where a processing operation will be unfair and reference can be made to those that result in direct discrimination as an example. However in practice, the lines to be drawn are far more blurred as the determination of what is fair is open to interpretation, at least until there is a Court ruling. Without Court of Justice rulings on the sticky issues running to the core of what fairness means in concrete contexts, there will always be uncertainty in the DPIA process in terms of the appropriate balance to be struck. There is a need to explore more deeply the relationship between fairness as a ‘core’ principle of the GDPR and the methodological approaches to the process of conducting a DPIA. Such an analysis would allow for a better understanding of the relationship between the principles provided for in Article 5 but also the various meanings attributed to fairness.

Here reference can also be made to \citet{butterworth2018ico} who argues that the ICO's focus on fairness in relation to artificial intelligence and machine learning seems to stretch the GDPR and its fair processing requirement to address the challenges such as collective harms. Building on this point, \citet{butterworth2018ico} finds that there may be a need for ‘legislation defining socially acceptable limits and controls on the application of artificial intelligence, and providing effective rights of redress for individuals and groups that may suffer harm.’

Indeed, it has repeatedly been suggested that, for example, consumer law can act as a toolbox for the mobilization of the protection of consumers in order to facilitate more holistic protection \cite{clifford2019pre,helberger2017perfect}. Such a development would allow legal protections to move beyond ‘the exclusive realm of informational privacy and self-determination’ \cite{koops2013decision}. This approach may also cater for the difficulties associated with trying to operationalize collective harms. That is, instead of focusing on a reconceptualization of how a group may fit within a fundamental rights framework, legislation may be adopted on the basis of collective concerns in the pursuit of human dignity, individual autonomy and personality in order to mitigate the negative effects of such developments. With this in mind, the recent publication of the draft of a Proposal for a Regulation of the European Parliament and of the Council laying down harmonized rules on Artificial Intelligence (Artificial Intelligence Act) and amending certain Union legislation acts, COM(2021) 206 final 2021/0106 (COD) is a clear indication that European Union policy makers appear to move in this direction as demonstrated by its proposed banning of certain Artificial Intelligence technologies and applications.

Such an approach seems to at least in part recognizes the limitations associated with the emphasis on controller accountability and the GDPR's decentered regulatory approach that is illustrative of (1) the focus on risk and responsiveness and (2) the enhanced focus on accountability and the auditing of performance \cite{clifford2018data}. Indeed, these concerns have a clear impact on the operation of the fairness principle due to the fact that there is an inherent reliance on commercial entities to take fairness considerations into account. It should be noted, however, that our discussion does not negate the usefulness of the DPIA as a process but rather recognizes its limitations and that of the fairness principle to truly cater for developments in machine learning in a comprehensive manner.  

Finally, our discussion does not render fairness metrics unsuitable for adoption within the requirement to conduct a DPIA. Controllers should be encouraged, and indeed are required, to consider the consequences of their personal data processing operations. Therefore, the DPIA process should be considered as an important element in the accountability principle linked trail established in the Regulation. Instead, it is suggested that the fairness principle in the context of DPIAs is unlikely in itself even if operationalized through fairness metrics (keeping in mind their limitations) to fully cater for the concerns associated with the development of automated decision making systems.

\section{Conclusion}

The data protection fairness principle plays an important role in the requirement to conduct a DPIA and the operation of this process. Fairness however, is rarely mentioned in the literature exploring the requirement to conduct a DPIA. Hence, there is a clear need for further research exploring the reasons for this omission more thoroughly and also in analyzing how this could be incorporated in the guidance issued by data protection authorities. As fairness is a core principle, it would be counter-intuitive to suggest that it plays no role in the determination of the potential impact of future processing operations. Given the rather nebulous nature of the fairness principle, this paper has explored the potential for fairness metrics to operationalize the principle in order to more adequately respond to the potential for unfair outcomes. Although there is certainly a role for fairness metrics in rendering the requirement to process personal data fairly more tangible, we have argued that such an approach also has significant limitations. Indeed, reference here can be made to (1) the technical challenges that have an impact on the usefulness of DPIAs irrespective of a controller’s willingness to actively engage in the process, (2) the context dependent nature of the fairness principle and the narrow but also varying interpretations of fairness according to different fairness metrics, and (3) the key role played by data controllers in the determination of what is fair. Hence, although fairness is key to the operation of DPIAs it is unlikely to cater for all our concerns related to the processing of personal data in particular in the context of the employment and deployment of Artificial Intelligence. As such, the arguments in this paper justify the need for a  more  open debate regarding the  role of the fairness principle, DPIAs, and the intersection of data protection with different policy agendas in the determination of what processing operations should be deemed de facto unfair. Our paper has therefore laid the foundation for a more detailed analysis of this topic in light of the forthcoming moves by, for instance, European Union policy makers to regulate Artificial Intelligence.

\section{Acknowledgments}

This project was supported by the Humanizing Machine Intelligence Grand Challenge at the Australian National University.

\bibliographystyle{ACM-Reference-Format}
\bibliography{sample-base}


\begin{thebibliography}{39}


\ifx \showCODEN    \undefined \def \showCODEN     #1{\unskip}     \fi
\ifx \showDOI      \undefined \def \showDOI       #1{#1}\fi
\ifx \showISBNx    \undefined \def \showISBNx     #1{\unskip}     \fi
\ifx \showISBNxiii \undefined \def \showISBNxiii  #1{\unskip}     \fi
\ifx \showISSN     \undefined \def \showISSN      #1{\unskip}     \fi
\ifx \showLCCN     \undefined \def \showLCCN      #1{\unskip}     \fi
\ifx \shownote     \undefined \def \shownote      #1{#1}          \fi
\ifx \showarticletitle \undefined \def \showarticletitle #1{#1}   \fi
\ifx \showURL      \undefined \def \showURL       {\relax}        \fi
\providecommand\bibfield[2]{#2}
\providecommand\bibinfo[2]{#2}
\providecommand\natexlab[1]{#1}
\providecommand\showeprint[2][]{arXiv:#2}

\bibitem[\protect\citeauthoryear{A29WP}{A29WP}{2017}]%
        {DPIA17}
\bibfield{author}{\bibinfo{person}{A29WP}.} \bibinfo{year}{2017}\natexlab{}.
\newblock \showarticletitle{Article 29 Working Party, Guidelines on Data
  Protection Impact Assessment (DPIA) and Determining Whether Processing Is
  “Likely to Result in a High Risk” for the Purposes of Regulation 2016/679
  (No WP248 rev.01, 4 October 2017) 1}.
\newblock  (\bibinfo{year}{2017}).
\newblock


\bibitem[\protect\citeauthoryear{Angwin, Jeff, Surya, and Lauren}{Angwin
  et~al\mbox{.}}{2016}]%
        {julia2016machine}
\bibfield{author}{\bibinfo{person}{Julia Angwin}, \bibinfo{person}{Larson
  Jeff}, \bibinfo{person}{Mattu Surya}, {and} \bibinfo{person}{Kirchner
  Lauren}.} \bibinfo{year}{2016}\natexlab{}.
\newblock \showarticletitle{Machine Bias: There's Software Used Across the
  Country to Predict Future Criminals and It's Biased Against Blacks}.
\newblock \bibinfo{journal}{\emph{ProPublica}} (\bibinfo{year}{2016}).
\newblock
\newblock
\shownote{\url{https://www.propublica.org/article/machine-bias-risk-assessments-in-criminal-sentencing.}}


\bibitem[\protect\citeauthoryear{Berk, Heidari, Jabbari, Kearns, and Roth}{Berk
  et~al\mbox{.}}{2018}]%
        {berk2018fairness}
\bibfield{author}{\bibinfo{person}{Richard Berk}, \bibinfo{person}{Hoda
  Heidari}, \bibinfo{person}{Shahin Jabbari}, \bibinfo{person}{Michael Kearns},
  {and} \bibinfo{person}{Aaron Roth}.} \bibinfo{year}{2018}\natexlab{}.
\newblock \showarticletitle{Fairness in criminal justice risk assessments: The
  state of the art}.
\newblock \bibinfo{journal}{\emph{Sociological Methods \& Research}}
  (\bibinfo{year}{2018}), \bibinfo{pages}{0049124118782533}.
\newblock


\bibitem[\protect\citeauthoryear{Buolamwini and Gebru}{Buolamwini and
  Gebru}{2018}]%
        {buolamwini2018gender}
\bibfield{author}{\bibinfo{person}{Joy Buolamwini} {and}
  \bibinfo{person}{Timnit Gebru}.} \bibinfo{year}{2018}\natexlab{}.
\newblock \showarticletitle{Gender shades: {I}ntersectional accuracy
  disparities in commercial gender classification}. In
  \bibinfo{booktitle}{\emph{Conference on Fairness, Accountability and
  Transparency}}. \bibinfo{pages}{77--91}.
\newblock


\bibitem[\protect\citeauthoryear{Butterworth}{Butterworth}{2018}]%
        {butterworth2018ico}
\bibfield{author}{\bibinfo{person}{Michael Butterworth}.}
  \bibinfo{year}{2018}\natexlab{}.
\newblock \showarticletitle{The ICO and artificial intelligence: The role of
  fairness in the GDPR framework}.
\newblock \bibinfo{journal}{\emph{Computer Law \& Security Review}}
  \bibinfo{volume}{34}, \bibinfo{number}{2} (\bibinfo{year}{2018}),
  \bibinfo{pages}{257--268}.
\newblock


\bibitem[\protect\citeauthoryear{Bygrave}{Bygrave}{2002}]%
        {bygrave2002data}
\bibfield{author}{\bibinfo{person}{LA Bygrave}.}
  \bibinfo{year}{2002}\natexlab{}.
\newblock \showarticletitle{Data protection law: approaching its rationale,
  logic and limits,(Vol. 10)}.
\newblock \bibinfo{journal}{\emph{Information Law Series. The Hague: Kluwer Law
  International}} (\bibinfo{year}{2002}).
\newblock


\bibitem[\protect\citeauthoryear{Chouldechova}{Chouldechova}{2017}]%
        {chouldechova2017fair}
\bibfield{author}{\bibinfo{person}{Alexandra Chouldechova}.}
  \bibinfo{year}{2017}\natexlab{}.
\newblock \showarticletitle{Fair prediction with disparate impact: A study of
  bias in recidivism prediction instruments}.
\newblock \bibinfo{journal}{\emph{Big data}} \bibinfo{volume}{5},
  \bibinfo{number}{2} (\bibinfo{year}{2017}), \bibinfo{pages}{153--163}.
\newblock


\bibitem[\protect\citeauthoryear{Christin, Rosenblat, and Boyd}{Christin
  et~al\mbox{.}}{2015}]%
        {christin2015courts}
\bibfield{author}{\bibinfo{person}{Ang{\`e}le Christin}, \bibinfo{person}{Alex
  Rosenblat}, {and} \bibinfo{person}{Danah Boyd}.}
  \bibinfo{year}{2015}\natexlab{}.
\newblock \showarticletitle{Courts and predictive algorithms}.
\newblock \bibinfo{journal}{\emph{Data \& Civil Right: Criminal Justice and
  Civil Rights Primer}} (\bibinfo{year}{2015}).
\newblock


\bibitem[\protect\citeauthoryear{Clifford and Ausloos}{Clifford and
  Ausloos}{2018}]%
        {clifford2018data}
\bibfield{author}{\bibinfo{person}{Damian Clifford} {and} \bibinfo{person}{Jef
  Ausloos}.} \bibinfo{year}{2018}\natexlab{}.
\newblock \showarticletitle{Data protection and the role of fairness}.
\newblock \bibinfo{journal}{\emph{Yearbook of European Law}}
  \bibinfo{volume}{37} (\bibinfo{year}{2018}), \bibinfo{pages}{130--187}.
\newblock


\bibitem[\protect\citeauthoryear{Clifford, Graef, and Valcke}{Clifford
  et~al\mbox{.}}{2019}]%
        {clifford2019pre}
\bibfield{author}{\bibinfo{person}{Damian Clifford}, \bibinfo{person}{Inge
  Graef}, {and} \bibinfo{person}{Peggy Valcke}.}
  \bibinfo{year}{2019}\natexlab{}.
\newblock \showarticletitle{Pre-formulated Declarations of Data Subject
  Consent—Citizen-Consumer Empowerment and the Alignment of Data, Consumer
  and Competition Law Protections}.
\newblock \bibinfo{journal}{\emph{German Law Journal}} \bibinfo{volume}{20},
  \bibinfo{number}{5} (\bibinfo{year}{2019}), \bibinfo{pages}{679--721}.
\newblock


\bibitem[\protect\citeauthoryear{CNIL}{CNIL}{2018}]%
        {ref12}
\bibfield{author}{\bibinfo{person}{CNIL}.} \bibinfo{year}{2018}\natexlab{}.
\newblock \showarticletitle{Commission Nationale Informatique \& Libertés,
  Privacy Impact Assessment Application to IOT Devices}.
\newblock  (\bibinfo{year}{2018}).
\newblock


\bibitem[\protect\citeauthoryear{Corbett-Davies, Pierson, Feller, Goel, and
  Huq}{Corbett-Davies et~al\mbox{.}}{2017}]%
        {corbett2017algorithmic}
\bibfield{author}{\bibinfo{person}{Sam Corbett-Davies}, \bibinfo{person}{Emma
  Pierson}, \bibinfo{person}{Avi Feller}, \bibinfo{person}{Sharad Goel}, {and}
  \bibinfo{person}{Aziz Huq}.} \bibinfo{year}{2017}\natexlab{}.
\newblock \showarticletitle{Algorithmic decision making and the cost of
  fairness}. In \bibinfo{booktitle}{\emph{Proceedings of the 23rd acm sigkdd
  international conference on knowledge discovery and data mining}}.
  \bibinfo{pages}{797--806}.
\newblock


\bibitem[\protect\citeauthoryear{DPC}{DPC}{2019}]%
        {ref11}
\bibfield{author}{\bibinfo{person}{DPC}.} \bibinfo{year}{2019}\natexlab{}.
\newblock \showarticletitle{Data Protection Commission, Guidance Note: Guide to
  Data Protection Impact Assessments (DPIAs)}.
\newblock  (\bibinfo{year}{2019}).
\newblock


\bibitem[\protect\citeauthoryear{Dwork, Hardt, Pitassi, Reingold, and
  Zemel}{Dwork et~al\mbox{.}}{2012}]%
        {dwork2012fairness}
\bibfield{author}{\bibinfo{person}{Cynthia Dwork}, \bibinfo{person}{Moritz
  Hardt}, \bibinfo{person}{Toniann Pitassi}, \bibinfo{person}{Omer Reingold},
  {and} \bibinfo{person}{Richard Zemel}.} \bibinfo{year}{2012}\natexlab{}.
\newblock \showarticletitle{Fairness through awareness}. In
  \bibinfo{booktitle}{\emph{Proceedings of the 3rd innovations in theoretical
  computer science conference}}. \bibinfo{pages}{214--226}.
\newblock


\bibitem[\protect\citeauthoryear{EDPB}{EDPB}{2020}]%
        {EDPBDPbD}
\bibfield{author}{\bibinfo{person}{EDPB}.} \bibinfo{year}{2020}\natexlab{}.
\newblock \showarticletitle{European Data Protection Board, Guidelines 4/2019
  on Article 25 Data Protection by Design and by Default (Adopted on 20 October
  2020) 1}.
\newblock  (\bibinfo{year}{2020}).
\newblock


\bibitem[\protect\citeauthoryear{EDPS}{EDPS}{2014}]%
        {european2014privacy}
\bibfield{author}{\bibinfo{person}{EDPS}.} \bibinfo{year}{2014}\natexlab{}.
\newblock \bibinfo{booktitle}{\emph{European Data Protection Supervisor,
  Privacy and Competitiveness in the Age of Big Data: The Interplay Between
  Data Protection, Competition Law and Consumer Protection in the Digital
  Economy}}.
\newblock \bibinfo{publisher}{European Data Protection Supervisor}.
\newblock


\bibitem[\protect\citeauthoryear{EDPS}{EDPS}{2020}]%
        {ref10}
\bibfield{author}{\bibinfo{person}{EDPS}.} \bibinfo{year}{2020}\natexlab{}.
\newblock \showarticletitle{European Data Protection Supervisor, Survey on Data
  Protection Impact Assessments under Article 39 of the Regulation (case
  2020-0066, 6 July 2020 1)}.
\newblock  (\bibinfo{year}{2020}).
\newblock


\bibitem[\protect\citeauthoryear{Friedler, Scheidegger, Venkatasubramanian,
  Choudhary, Hamilton, and Roth}{Friedler et~al\mbox{.}}{2019}]%
        {friedler2019comparative}
\bibfield{author}{\bibinfo{person}{Sorelle~A Friedler}, \bibinfo{person}{Carlos
  Scheidegger}, \bibinfo{person}{Suresh Venkatasubramanian},
  \bibinfo{person}{Sonam Choudhary}, \bibinfo{person}{Evan~P Hamilton}, {and}
  \bibinfo{person}{Derek Roth}.} \bibinfo{year}{2019}\natexlab{}.
\newblock \showarticletitle{A comparative study of fairness-enhancing
  interventions in machine learning}. In \bibinfo{booktitle}{\emph{Proceedings
  of the conference on fairness, accountability, and transparency}}.
  \bibinfo{pages}{329--338}.
\newblock


\bibitem[\protect\citeauthoryear{Hacker}{Hacker}{2018}]%
        {hacker2018teaching}
\bibfield{author}{\bibinfo{person}{Philipp Hacker}.}
  \bibinfo{year}{2018}\natexlab{}.
\newblock \showarticletitle{Teaching fairness to artificial intelligence:
  Existing and novel strategies against algorithmic discrimination under EU
  law}.
\newblock  (\bibinfo{year}{2018}).
\newblock


\bibitem[\protect\citeauthoryear{Hardt, Price, and Srebro}{Hardt
  et~al\mbox{.}}{2016}]%
        {hardt2016equality}
\bibfield{author}{\bibinfo{person}{Moritz Hardt}, \bibinfo{person}{Eric Price},
  {and} \bibinfo{person}{Nati Srebro}.} \bibinfo{year}{2016}\natexlab{}.
\newblock \showarticletitle{Equality of opportunity in supervised learning}. In
  \bibinfo{booktitle}{\emph{Advances in neural information processing
  systems}}. \bibinfo{pages}{3315--3323}.
\newblock


\bibitem[\protect\citeauthoryear{Helberger, Zuiderveen~Borgesius, and
  Reyna}{Helberger et~al\mbox{.}}{2017}]%
        {helberger2017perfect}
\bibfield{author}{\bibinfo{person}{Natali Helberger}, \bibinfo{person}{Frederik
  Zuiderveen~Borgesius}, {and} \bibinfo{person}{Agustin Reyna}.}
  \bibinfo{year}{2017}\natexlab{}.
\newblock \showarticletitle{The perfect match? A closer look at the
  relationship between EU consumer law and data protection law}.
\newblock \bibinfo{journal}{\emph{Common Market Law Review}}
  \bibinfo{volume}{54}, \bibinfo{number}{5} (\bibinfo{year}{2017}).
\newblock


\bibitem[\protect\citeauthoryear{ICO}{ICO}{2018}]%
        {ref13}
\bibfield{author}{\bibinfo{person}{ICO}.} \bibinfo{year}{2018}\natexlab{}.
\newblock \showarticletitle{Information Commissioner’s Office, Sample DPIA
  Template (No20180209v0.3)}.
\newblock  (\bibinfo{year}{2018}).
\newblock


\bibitem[\protect\citeauthoryear{ICO}{ICO}{2020}]%
        {ICO2020}
\bibfield{author}{\bibinfo{person}{ICO}.} \bibinfo{year}{2020}\natexlab{}.
\newblock \showarticletitle{Information Commissioner’s Office, Guidance on AI
  and data protection (20203006 0.0.39)}.
\newblock \bibinfo{journal}{\emph{ICO}} (\bibinfo{year}{2020}).
\newblock


\bibitem[\protect\citeauthoryear{ICO}{ICO}{2021}]%
        {ICO2021}
\bibfield{author}{\bibinfo{person}{ICO}.} \bibinfo{year}{2021}\natexlab{}.
\newblock \showarticletitle{Information Commissioner’s Office, Guidance on
  Data Protection Impact Assessments (DPIAs) (No 2021101 1.1.64)}.
\newblock  (\bibinfo{year}{2021}).
\newblock


\bibitem[\protect\citeauthoryear{Janssen}{Janssen}{2020}]%
        {janssen2020approach}
\bibfield{author}{\bibinfo{person}{Heleen Janssen}.}
  \bibinfo{year}{2020}\natexlab{}.
\newblock \showarticletitle{An approach for a fundamental rights impact
  assessment to automated decision-making}.
\newblock \bibinfo{journal}{\emph{International Data Privacy Law}}
  \bibinfo{volume}{10} (\bibinfo{year}{2020}).
\newblock


\bibitem[\protect\citeauthoryear{Kasirzadeh and Smart}{Kasirzadeh and
  Smart}{2021}]%
        {kasirzadeh2021use}
\bibfield{author}{\bibinfo{person}{Atoosa Kasirzadeh} {and}
  \bibinfo{person}{Andrew Smart}.} \bibinfo{year}{2021}\natexlab{}.
\newblock \showarticletitle{The use and misuse of counterfactuals in ethical
  machine learning}. In \bibinfo{booktitle}{\emph{Proceedings of the 2021 ACM
  Conference on Fairness, Accountability, and Transparency}}.
  \bibinfo{pages}{228--236}.
\newblock


\bibitem[\protect\citeauthoryear{Khandani, Kim, and Lo}{Khandani
  et~al\mbox{.}}{2010}]%
        {khandani2010consumer}
\bibfield{author}{\bibinfo{person}{Amir~E Khandani}, \bibinfo{person}{Adlar~J
  Kim}, {and} \bibinfo{person}{Andrew~W Lo}.} \bibinfo{year}{2010}\natexlab{}.
\newblock \showarticletitle{Consumer credit-risk models via machine-learning
  algorithms}.
\newblock \bibinfo{journal}{\emph{Journal of Banking \& Finance}}
  \bibinfo{volume}{34}, \bibinfo{number}{11} (\bibinfo{year}{2010}),
  \bibinfo{pages}{2767--2787}.
\newblock


\bibitem[\protect\citeauthoryear{Kleinberg}{Kleinberg}{2018}]%
        {kleinberg2018inherent}
\bibfield{author}{\bibinfo{person}{Jon Kleinberg}.}
  \bibinfo{year}{2018}\natexlab{}.
\newblock \showarticletitle{Inherent trade-offs in algorithmic fairness}. In
  \bibinfo{booktitle}{\emph{Abstracts of the 2018 ACM International Conference
  on Measurement and Modeling of Computer Systems}}. \bibinfo{pages}{40--40}.
\newblock


\bibitem[\protect\citeauthoryear{Koops}{Koops}{2013}]%
        {koops2013decision}
\bibfield{author}{\bibinfo{person}{Bert-Jaap Koops}.}
  \bibinfo{year}{2013}\natexlab{}.
\newblock \showarticletitle{On decision transparency, or how to enhance data
  protection after the computational turn}.
\newblock \bibinfo{journal}{\emph{Privacy, due process and the computational
  turn: the philosophy of law meets the philosophy of technology}}
  (\bibinfo{year}{2013}), \bibinfo{pages}{189--213}.
\newblock


\bibitem[\protect\citeauthoryear{Kosta}{Kosta}{2020}]%
        {kosta20}
\bibfield{author}{\bibinfo{person}{Elini Kosta}.}
  \bibinfo{year}{2020}\natexlab{}.
\newblock \showarticletitle{Artilce 35. Data Protection Impact Assessment}. In
  \bibinfo{booktitle}{\emph{The EU General Data ProtectionRegulation (GDPR): A
  Commentary}} \emph{(\bibinfo{series}{Christopher Kuner et al.})}.
\newblock


\bibitem[\protect\citeauthoryear{Kusner, Loftus, Russell, and Silva}{Kusner
  et~al\mbox{.}}{2017}]%
        {kusner2017counterfactual}
\bibfield{author}{\bibinfo{person}{Matt~J Kusner}, \bibinfo{person}{Joshua
  Loftus}, \bibinfo{person}{Chris Russell}, {and} \bibinfo{person}{Ricardo
  Silva}.} \bibinfo{year}{2017}\natexlab{}.
\newblock \showarticletitle{Counterfactual fairness}. In
  \bibinfo{booktitle}{\emph{Advances in Neural Information Processing
  Systems}}. \bibinfo{pages}{4066--4076}.
\newblock


\bibitem[\protect\citeauthoryear{Lynskey}{Lynskey}{2014}]%
        {lynskey2014deconstructing}
\bibfield{author}{\bibinfo{person}{Orla Lynskey}.}
  \bibinfo{year}{2014}\natexlab{}.
\newblock \showarticletitle{Deconstructing data protection: the
  ‘added-value’of a right to data protection in the EU legal order}.
\newblock \bibinfo{journal}{\emph{International \& Comparative Law Quarterly}}
  \bibinfo{volume}{63}, \bibinfo{number}{3} (\bibinfo{year}{2014}),
  \bibinfo{pages}{569--597}.
\newblock


\bibitem[\protect\citeauthoryear{Malgieri}{Malgieri}{2020}]%
        {malgieri2020concept}
\bibfield{author}{\bibinfo{person}{Gianclaudio Malgieri}.}
  \bibinfo{year}{2020}\natexlab{}.
\newblock \showarticletitle{The concept of fairness in the GDPR: a linguistic
  and contextual interpretation}. In \bibinfo{booktitle}{\emph{Proceedings of
  the 2020 Conference on Fairness, Accountability, and Transparency}}.
  \bibinfo{pages}{154--166}.
\newblock


\bibitem[\protect\citeauthoryear{Mehrabi, Morstatter, Saxena, Lerman, and
  Galstyan}{Mehrabi et~al\mbox{.}}{2019}]%
        {mehrabi2019survey}
\bibfield{author}{\bibinfo{person}{Ninareh Mehrabi}, \bibinfo{person}{Fred
  Morstatter}, \bibinfo{person}{Nripsuta Saxena}, \bibinfo{person}{Kristina
  Lerman}, {and} \bibinfo{person}{Aram Galstyan}.}
  \bibinfo{year}{2019}\natexlab{}.
\newblock \showarticletitle{A survey on bias and fairness in machine learning}.
\newblock \bibinfo{journal}{\emph{arXiv preprint arXiv:1908.09635}}
  (\bibinfo{year}{2019}).
\newblock


\bibitem[\protect\citeauthoryear{Nemitz}{Nemitz}{2018}]%
        {nemitz2018constitutional}
\bibfield{author}{\bibinfo{person}{Paul Nemitz}.}
  \bibinfo{year}{2018}\natexlab{}.
\newblock \showarticletitle{Constitutional democracy and technology in the age
  of artificial intelligence}.
\newblock \bibinfo{journal}{\emph{Philosophical Transactions of the Royal
  Society A: Mathematical, Physical and Engineering Sciences}}
  \bibinfo{volume}{376}, \bibinfo{number}{2133} (\bibinfo{year}{2018}),
  \bibinfo{pages}{20180089}.
\newblock


\bibitem[\protect\citeauthoryear{Obermeyer, Powers, Vogeli, and
  Mullainathan}{Obermeyer et~al\mbox{.}}{2019}]%
        {obermeyer2019dissecting}
\bibfield{author}{\bibinfo{person}{Ziad Obermeyer}, \bibinfo{person}{Brian
  Powers}, \bibinfo{person}{Christine Vogeli}, {and} \bibinfo{person}{Sendhil
  Mullainathan}.} \bibinfo{year}{2019}\natexlab{}.
\newblock \showarticletitle{Dissecting racial bias in an algorithm used to
  manage the health of populations}.
\newblock \bibinfo{journal}{\emph{Science}} \bibinfo{volume}{366},
  \bibinfo{number}{6464} (\bibinfo{year}{2019}), \bibinfo{pages}{447--453}.
\newblock


\bibitem[\protect\citeauthoryear{Office}{Office}{2017}]%
        {ICO2017}
\bibfield{author}{\bibinfo{person}{Information~Commissioner’s Office}.}
  \bibinfo{year}{2017}\natexlab{}.
\newblock \showarticletitle{Big Data, Artificial Intelligence, Machine Learning
  and Data Protection (No 20170904)}.
\newblock \bibinfo{journal}{\emph{Version: 2.2}} (\bibinfo{year}{2017}).
\newblock


\bibitem[\protect\citeauthoryear{Pleiss, Raghavan, Wu, Kleinberg, and
  Weinberger}{Pleiss et~al\mbox{.}}{2017}]%
        {pleiss2017fairness}
\bibfield{author}{\bibinfo{person}{Geoff Pleiss}, \bibinfo{person}{Manish
  Raghavan}, \bibinfo{person}{Felix Wu}, \bibinfo{person}{Jon Kleinberg}, {and}
  \bibinfo{person}{Kilian~Q Weinberger}.} \bibinfo{year}{2017}\natexlab{}.
\newblock \showarticletitle{On fairness and calibration}. In
  \bibinfo{booktitle}{\emph{Advances in Neural Information Processing
  Systems}}. \bibinfo{pages}{5680--5689}.
\newblock


\bibitem[\protect\citeauthoryear{Yeung}{Yeung}{2018}]%
        {yeung2018study}
\bibfield{author}{\bibinfo{person}{Karen Yeung}.}
  \bibinfo{year}{2018}\natexlab{}.
\newblock \showarticletitle{A study of the implications of advanced digital
  technologies (including AI systems) for the concept of responsibility within
  a human rights framework}.
\newblock \bibinfo{journal}{\emph{Council of Europe) I-AUT2018) 05 29}}
  (\bibinfo{year}{2018}).
\newblock


\end{thebibliography}

\end{document}